# A New 2D Energy Balance Model For Simulating the Climates of Rapidly- and Slowly-Rotating Terrestrial Planets

Ramses M. Ramirez[1],

## Abstract

Energy balance models (EBMs), alongside radiative-convective climate models (RCMs) and global climate models (GCMs), are useful tools for simulating planetary climates. Historically, planetary and exoplanetary EBMs have solely been 1D latitudinally-dependent models with no longitudinal dependence, until the study of Okuya et al., which focused on simulating synchronously-rotating planets. Following the work of Okuya et al., I have designed the first 2D EBM (PlaHab) that can simulate $N_2$-$CO_2$-$H_2O$-$H_2$ atmospheres of both rapidly-rotating and synchronously-rotating planets, including Mars, Earth, and exoplanets located within their circumstellar habitable zones. PlaHab includes physics for both water and $CO_2$ condensation. Regional topography can be incorporated. Here, I have specifically applied PlaHab to investigate present Earth, early Mars, TRAPPIST-1e and Proxima Centauri b, representing examples of habitable (and potentially habitable) worlds in our solar system and beyond. I compare my EBM results against those of other 1D and 3D models, including those of the recent Trappist-1 Habitable Atmosphere (THAI) comparison project. Overall, EBM results are consistent with those of other 1D and 3D models although inconsistencies among all models continue to be related to the treatment of clouds and other known differences between EBMs and GCMs, including heat transport parameterizations. Although two-dimensional EBMs are a relatively new entry in the study of planetary/exoplanetary climates, their ease-of-use, speed, flexibility, wide applicability, and greater complexity (relative to 1D models), may indicate an ideal combination for the modeling of planetary and exoplanetary atmospheres alike.

1     University of Central Florida, Department of Physics, Planetary Sciences Group, Orlando, Fl. 32816, USA

---

1    Corresponding author:  Ramses.Ramirez@ucf.edu

# 1. INTRODUCTION

Energy balance models (EBMs), like their radiative-convective and 3D global climate model cousins, have been developed to study the climate of Earth, specifically the ice-albedo feedback in the case of EBMs. The original EBMs, independently produced by Sellers (1969) and Budyko (1969), equated the incoming solar radiation and the outgoing longwave radiation for energy balance in planetary atmospheres, a standard practice today. These Budyko-Seller models and subsequent variants (e.g. North and Coakley 79; 81) consider mean annual surface temperatures in latitudinal zones. Notably, both Budyko (1969) and Sellers (1969) predicted that the Earth would transition to a water ice-covered (snowball) planet should the solar radiation decrease by only a few percent. Held and Suarez (1974) improved upon the Budyko formula whereas North (1975ab) used a constant thermal diffusion coefficient (as is commonly done today) to solve the Budyko/Sellers Earth water ice cap model. Earth habitability EBMs have also been developed to study the effects of high obliquity (Williams and Kasting 1997).

The success of EBMs in simulating the Earth's climate (see North et al. 1981 for a review), along with their simplicity, have since spawned the development of EBMs to investigate our neighboring planets and beyond. In our solar system, EBMs have now studied the climates of both present and early Mars (e.g. Armstrong et al. 2004; Savijarvi 2004; Batalha et al. 2016; Hayworth et al. 2020). Exoplanetary applications have been particularly numerous and include (but are not limited to) studies that have assessed the variation of planetary/orbital parameters on planetary habitability (e.g. Spiegel et al. 2008; 2009; 2010; Dressing et al. 2010; Armstrong et al. 2014), evaluated the strength of the $CO_2$ ice-albedo feedback for both warm and cold start planets(e.g. Kadoya and Tajika 2014;2016; Haqq-Misra et al. 2016; 2019; Bonati and Ramirez 2021), water ice coverage in habitable zone planets (Wilhelm et al. 2022),water worlds (Ramirez and Levi 2018), Milankovitch cycles in planets orbiting binary stellar systems (e.g. Forgan 2014, Quarles 2022), exomoons (Williams and Kasting 1997, Forgan and Dobos 2016), the habitable zone for complex life (Ramirez 2020a) and the impact of atmospheric pressure on habitable zone boundaries(e.g. Vladilo et al. 2013; 2015; Ramirez 2020b).

The use of 1D latitudinally-dependent climate models is the common link in the aforementioned studies (including the terrestrial ones), which may suffice to model many cases of rapidly-rotating planets with effective lateral heat transport. However, 1D EBMs are generally ill-equipped to simulate the climates of planets with large longitudinal temperature variations, such as synchronously-rotating exoplanets. Nevertheless, 1D EBMs can partially address this issue by employing a coordinate transformation that enables simulation of tidally-locked planets (Checlair 2017; Haqq-Misra and Hayworth 2022). However, even with such modifications, key longitudinal information (e.g. surface temperatures, albedo, fluxes) remains inaccessible in such models, which partially limits their application and ability to compare against higher order codes, including 3D GCMs. For instance, it is much easier to apply different cloud models for the day- and nightsides or incorporate regionally-varying topography with 2D EBMs and GCMs than by utilizing their 1D counterparts. One-dimensional models are also unable to produce important astronomical observables that require multiple dimensions, including the generation of disc-integrated spectra (e.g. Fauchez et al. 2022).

The study of Okuya et al. (2019), which assessed the habitability of synchronously-rotating planets (with a focus on binary star systems), is the first 2D exoplanetary EBM. Okuya et al. (2019) developed their model by expanding the energy balance equation to two dimensions and by adding stellar flux longitudinal variations, implementing ideas from 2D EBMs used to model the Earth (e.g. North and Coakley 1979; Pollard et al. 1983). Their diffusion parameterization was rather simple, only incorporating pressure, gravity, and atmospheric heat capacity (ignoring other factors like rotation rate and molecular mass, e.g. Williams and Kasting 1997), only scaling appropriately to generate reasonable temperature gradients. Their model did not explicitly model $CO_2$ condensation mass losses and gains (only finding the condensation temperature). Cloud albedo was given a constant value. In spite of such approximations, it was able to reasonably reproduce the latitudinal and longitudinal temperature structure for Proxima Centauri b (assuming an Earth-like atmospheric composition) as compared with the equivalent 3D GCM calculation (Turbet et al. 2016). Given that such simulations can be completed in a fraction of the time that a 3D GCM would take, Okuya et al. illustrated the great potential of 2D EBMs for planetary and exoplanetary science. For example, whereas a GCM may require a couple of days to simulate a particular atmosphere, a 2D EBM can perform the equivalent calculation in mere seconds. Thus, 2D EBMs are useful for exploring the sensitivity of results to a particular parameter which may require a large number of simulations that a GCM would take prohibitively long (months or even years) to accomplish the same feat.

Here, I've developed a new 2D **Pla**netary **Hab**itability energy balance climate model (PlaHab) that can simulate both rapidly- and slowly-rotating terrestrial solar system planets and exoplanets with $N_2$-$CO_2$-$H_2O$-$H_2$ atmospheres orbiting F – M main-sequence stars. The model is inspired by the work of Okuya et al. (2019) and earlier 1D models, such as Williams and Kasting (1997). However, here I showcase PlaHab's ability to reproduce the climates of rapidly-rotating planets such as present Earth and early Mars, and tidally-locked worlds (Proxima Centauri b and TRAPPIST-1e). The goal here was to produce a *fast* 2D EBM that self-consistently calculates the climates of these (and other) terrestrial worlds with virtually no further tuning or scaling of parameters outside of the initial universal inputs. I conclude with a model assessment and suggested future directions.

## 2. METHODS
### 2.1 Governing Equations

Our new 2D advanced EBM (PlaHab) is itself derived from a family of similar 1D latitudinally-dependent codes (e.g. North & Coakley 1979; North et al. 1981; Williams and Kasting 1997; Ramirez and Levi 2018; Bonati and Ramirez 2021). As with these and other 1D EBMs, PlaHab follows the radiative energy balance principle that planets in thermal equilibrium radiate as much energy to space as they receive from their host stars. The atmospheric-ocean-energy balance equation follows Okuya et al. (2019), which itself comes from earlier sources (e.g. North and Coakley 1979; Pollard 1983), except PlaHab also includes the effects of $CO_2$ condensation and sublimation (Equation. 1):

$$S(1-A) = C_p \frac{\partial T(\theta,\phi,t)}{\partial t} + I - \frac{1}{\cos\theta}\frac{\partial}{\partial\theta}D_1 \cos\theta \frac{\partial T(\theta,\phi,t)}{\partial\theta} - \frac{1}{\cos^2\theta}\frac{\partial}{\partial\phi}D_2 \frac{\partial T(\theta,\phi,t)}{\partial\phi} - L\left(\frac{dM_{col,CO_2}}{dt}\right) \quad \text{(Eqn. 1)}$$

where $C_p$ is the effective heat capacity, T is the temperature at a given grid point, $\theta$ is latitude, $\phi$ is the longitudinal coordinate, *t* is time, *I* is the infrared emission to space, *A* is the planetary albedo, *L* is the latent heat flux per unit mass of $CO_2$ (5.9x105 J/kg; Forget et al. 1998). $M_{col,CO2}$ is the column mass of $CO_2$ which sublimates or condenses on to the planetary surface, with $D_1$ and $D_2$ being the latitudinal and longitudinal diffusion coefficients, respectively. In EBMs, heat transport is parameterized as diffusion for simplicity. Equation 1 is solved at every time step using a second order finite differencing scheme.

The thermal diffusion coefficients $(D_o, D)$ are parameterized similarly to that in Williams and Kasting (1997)(Equations 2 and 3) except for PlaHab's inclusion of the longitudinal direction:

$$\left(\frac{D_1}{D_{o,1}}\right) = \left(\frac{P}{P_o}\right)\left(\frac{C_p}{C_{po}}\right)\left(\frac{m_o}{m}\right)^2 \left(\frac{\Omega_o}{\Omega}\right)^2 \left(\frac{FH_2O}{FH_2O_o}\right)^2 \quad \text{(Eqn. 2)}$$

$$\left(\frac{D_2}{D_{o,2}}\right) = \left(\frac{P}{P_o}\right)\left(\frac{C_p}{C_{po}}\right)\left(\frac{m_o}{m}\right)^2 \left(\frac{\Omega}{\Omega_o}\right)\left(\frac{FH_2O}{FH_2O_o}\right)^2 \quad \text{(Eqn. 3)}$$

where the subscripts 1 and 2 refer to latitude and longitude, respectively and the subscript "o" values represent those for Earth (except for the diffusion coefficients, see below). The terms are the heat capacity ($C_{po}$ = 10$^3$ g$^{-1}$kgK$^{-1}$), $P$ is the atmospheric pressure ($P_o$ = 1 bar), $m$ is the atmospheric molar mass ($m_o$ = 29 g/mol), $\Omega$ is the planetary rotation rate ($\Omega_o$ = 7.27x10$^{-5}$ r/s) and $FH_2O$ is a water vapor mixing ratio ($FH_2O_o$ = 0.0155) correction factor. The latter simulates the importance of latent heat release and transport at high temperatures by tracking the exponential dependence as predicted by the Clausius-Clapeyron equation(e.g. Caballero and Lagen 2005; Rose et al. 2017). Without this temperature dependence term, equator-to-pole transport is vastly underestimated at warm temperatures and greatly overestimated at cold temperatures as compared to GCM simulations(Bonati and Ramirez 2021). Bonati and Ramirez (2021) had crudely parameterized this term by deriving a transport dependence based on one GCM model's results (Turbet et al. 2017), which exhibited an exponential behavior similar to that of Clausius-Clapeyron. Here, I have directly applied Clausius-Clapeyron to capture this exponential increase in latent heat transport using self-consistent physics. I find that this yields model behavior that is similar, albeit not identical, to that obtained in Bonati and Ramirez (2021), while often producing results that are comparable to those obtained by dynamic ocean GCMs (Pollard 1983).

The values for the thermal diffusion depend on whether a planet is rapidly- or synchronously-rotating and whether water vapor is present. For Earth, $D_{o,1}$ = 0.5 Wm$^{-2}$K$^{-1}$, and $D_{o,2}$ = 0.02 Wm$^{-2}$K$^{-1}$, for $\theta$< 45 degrees ($D_{o,2}$0.0002 at higher latitudes). I use the same base coefficients for Mars (another rapid rotator). Following Okuya et al. (2019), I assume that the *calculated* thermal diffusion coefficients ($D_{(1,2)}$) for dry desert planets are half those of aqua planets(e.g. Pollard 1983; Oort and Vondar Haar 1976).

Although equations 2 and 3 appear similar in form, the rotation rate dependence in each is different. According to Equation 2, latitudinal (i.e. meridional) transport should intensify as the rotation rate decreases (i.e. Rossby Number increases). This is because the Hadley cells expand to higher latitudes in both hemispheres, increasing poleward transport which reduces equator-pole temperature contrast (e.g. Merlis and Schneider 2010; Kaspi and Showman 2015; Komacek and Abbot 2019). In contrast, the zonal jets that are prevalent at higher rotation rates weaken for slow rotators (e.g. Merlis and Schneider

2010; Kaspi and Showman 2015; Komacek and Abbot 2019), which implies a diminished longitudinal transport for those cases (Equation 3). Here, I assume a linear relationship between longitudinal ($D_2$) transport and rotation rate. I note that computing the exact proportionality between zonal and diffusive transport is not necessary because this is addressed by proper fitting of the $D_o$ terms.

Even though the basic physics behind equations 2 and 3 are sound, $D_{o,1}$ and $D_{o,2}$ represent tuning parameters that can further improve model agreement with both EBMs and observations. Thus, I ran a suite of PlaHab simulations at different rotation rates (i.e. 0.25 – 4x Earth's rotation rate), finding the $D_{o,1}$ and $D_{o,2}$ that was needed to achieve convergence in each case. The diffusion parameter solutions were then fitted to the following quadratic function (Equation 4):

$$D_o, (1,2) = a_1 \eta^2 + a_2 \eta + a_3 \qquad \text{(Eqn. 4)}$$

Here, $\eta = \Omega/\Omega_o$. More details on the curve fit are given in the Appendix to this article.

For rapidly-rotating planets, the stellar flux (diurnally- and globally-averaged) is as follows (Equation 5):

$$S = \frac{q_o}{\pi} \left(\frac{1.0 AU}{a}\right)^2 (H \sin\theta \sin\delta + \cos\theta \cos\delta \sin H) \qquad \text{(Eqn. 5)}$$

Here, H is the radian half-day length whereas $\delta$ is the declination angle. This is the same equation utilized by Williams and Kasting (1997) and other 1D EBMs (Haqq-Misra et al. 2016;2022; Bonati and Ramirez (2021). However, for synchronously-rotating planets, the incident stellar flux calculated for any region on the planet is computed as (Equation 6):

$$S = q_o \left(\frac{1.0 AU}{a}\right)^2 max(\cos\theta \cos\phi, 0.) \qquad \text{(Eqn. 6)}$$

This adapts the longitudinal dependence from Okuya et al. (2019), averaged over the zenith angle. This particular form of the equation is derived from Equation 5, assuming a zero obliquity which is a reasonable expectation following tidal dissipation (Okuya et al. 2019).

PlaHab is flexible and can adapt any number of latitudinal and longitudinal bands. For rapidly-rotating planets, I find that implementing 36 five-degree latitude bands and 12 thirty-degree longitude bands (480 total grid points) is sufficient to spatially resolve cloudy regions, surface ice regions, continents, and oceans, while maintaining reasonable computational costs (~2 - 20 seconds per simulation) and accuracy. However, I find that at least 20 longitude bands are needed to accurately model the planetary flux distribution for synchronously-rotating planets, which is the number of longitudinal bands I implement here for these worlds. By default, flat topography is nominally assumed for PlaHab, but it is possible to prescribe regional topography as I demonstrate below in the case of early Mars.

PlaHab's treatment of clouds is not radically different from that in 1D EBMs (e.g. Bonati and Ramirez 2021), although it has been expanded, based on similar parameterizations derived for one GCM (Xu & Krueger 1991). As the atmosphere warms near the freezing point of water (> 263 K), water clouds form, with the latitudinal coverage (c) dictated by (Equation 7):

$$c(\theta, \phi) = min\left(0.72 log\left(\frac{F_C}{F_E} + 1\right), 1\right) \quad \text{(Eqn. 7)}$$

Here, $F_C$ is the convective heat flux whereas $F_E$ is the corresponding value for Earth at 288 K (~90 W/m²). The convective heat flux is computed by adding the sensible $(F_S)$ and latent $(F_L)$ heat fluxes (Equations 8 and 9)(e.g. Kasting et al. 2014):

$$F_L = (1 - RH)LC_d v \rho_s q_{h_2o} \quad \text{(Eqn. 8)}$$

$$F_S = c_p \rho_s C_d v \Delta T \quad \text{(Eqn. 9)}$$

Here, $RH$ is the surface relative humidity, $L$ is the latent heat of vaporization, $C_d$ is the drag coefficient (1.5x10$^{-3}$; Hidy 1972; Pond et al. 1974), $v$ is velocity (10 m/s; Pond et al. 1974), $\rho$ is near-surface atmospheric density, $q_{h_2o}$ is specific humidity, $c_p$ is the specific heat and $\Delta T$ is the near-surface temperature difference (7x10$^{-3}$ K, which calibrates well with Earth). The velocity is consistent with suggestions that a thick early Martian atmosphere would have likely exhibited near-surface wind speeds similar to those of present Earth (i.e. Ramirez and Kasting 2017; Ramirez 2017).

This model predicts ~50% cloud cover for Earth at 288 K, increasing at higher temperatures. The version in PlaHab now has longitudinal and latitudinal dependencies.

I have also constructed a similar simple parameterization that represents the overall greenhouse effect of water vapor clouds (Equation 10):

$$cir(\theta, \phi) = min\left(-7log\left(\frac{F_C}{F_E} + 1\right), -10\right) \quad \text{(Eqn. 10)}$$

Here, cir is the cloud infrared flux (in W/m²), which is added to the overall thermal emission. This expression (Equation 10) scales with the convective heat fluxes and produces the right amount of cloud infrared greenhouse effect that yields an overall 288 K Earth mean surface temperature.

At large enough stellar fluxes (e.g. habitable zone), convective fluxes are high enough on the daysides of synchronously-rotating terrestrial planets to form highly-reflective clouds that stabilize the climate (Yang et al. 2013; 2014). I handle this by deriving the following dayside cloud albedo parameterization (cpalb), which contributes to the total top-of-atmosphere albedo values (Equation 11):

$$cpalb(\theta, \phi) = min\left(c_p log\left(\frac{F_C}{F_E} + 1\right), 1\right) \quad \text{(Eqn. 11)}$$

Here $c_p$ is the dayside cloud enhancement factor, which is parameterized in accordance with the mean surface temperature (see Appendix). This parameterization yields a similar cloud albedo distribution to that computed by a recent TRAPPIST-1e GCM comparison (Sergeev et al. 2022).

$CO_2$ clouds in the current version of the model are radiatively-inactive. This is probably not a major issue given that their greenhouse effect is likely small in dense $CO_2$ atmospheres (Kitzmann 2016). Even so, our $CO_2$ clouds impact the planetary albedo and cloud cover is found by tracing an adiabat to the point of first condensation (Williams and Kasting 97; Bonati and Ramirez 2021). All such $CO_2$ clouds are assumed to form in that single layer.

As with similar 1D EBMs (e.g. Ramirez and Levi 2018, Bonati and Ramirez 2021; Haqq-Misra and Hayworth 2023), PlaHab employs radiative transfer lookup tables that are directly computed by a single column radiative-convective climate (RCM) model (e.g. Kasting et al. 1993; Ramirez and Kaltenegger 2017;2018). That model subdivides atmospheres into multiple vertical log layers (I implement 101 here) that extend from the ground to the top of the atmosphere (assumed to be $1 \times 10^{-5}$ bar here. Thus, each PlaHab grid point represents the surface properties from 1 RCM column. A standard moist convective adjustment is used in the RCM (Manabe and Wetherald 1967). The radiative transfer employs 55 infrared and 38 solar bands, respectively. Stellar spectra for F – M stars are derived from the Bt-Settl database (e.g. Allard et al. 2003; 2007). The RCM model includes the most up-to-date $CO_2$-$CO_2$, $N_2$-$N_2$, $N_2$-$H_2$, $H_2$-$H_2$, and $CO_2$-$H_2$ collision-induced absorption (CIA) parameterizations (Gruzka and Borysow 1997; Borysow 1998; Borysow 2002; Baranov et al. 2004; Wordsworth et al. 2010; Richard et al. 2012). PlaHab employs a multi-log-linear interpolation scheme to calculate surface temperature (150 – 390 K), total atmospheric pressure ($N_2$, $CO_2$, $H_2$) (up to 100 bars), surface albedo and zenith angle to a very high degree of accuracy (<1% error, Vladilo et al. 2013;2015), far surpassing that of models that utilize polynomial fits, which often accrue errors exceeding 20% (Haqq-Misra et al. 2016).

The surface water vapor partial pressure at every model grid point is calculated via the Clausius-Clapeyron equation, assuming a surface relative humidity of 100% for the wet synchronously-rotating aquaplanet cases (see results section) and 0% for the dry planets (in which case the relative humidity is zero). In the case of Mars and Earth, an Earth-like 77% relative humidity was used for the surface (Manabe-Wetherald 1967), following Ramirez et al. (2020). However, on the real Earth (for instance), a 3D GCM would exhibit lateral surface relative humidity variations that are consistent with topographical differences. However, most of our cases employ flat topography and I find that the differences are minor enough such that the zonal/meridional temperature results PlaHab produces remain comparable to those that a GCM may produce.

Following previous work (e.g., Ramirez 2020ab; Bonati and Ramirez 2021), Fresnel data is implemented for ocean reflectance as a function of zenith angle. Ice absorption is a function of 2 channels (UV/VIS and near-infrared), with their relative contributions depending on star type (e.g. Shields et al. 2013).

At each grid point, I have implemented the following surface albedo (α) parameterization for snow/ice mixtures (e.g. Ramirez 2020ab; Bonati and Ramirez 2021), similar to that in Curry et al. (2001)(Equations 12 and 13).

$$\alpha(visible) = \begin{bmatrix} 0.7; & T \leq 263.15 \\ 0.7 - 0.020(T - 263.15); & 263.15 < T < 273.15 \\ 0.22; & T \geq 273.15 \end{bmatrix} \quad \text{(Eqn. 12)}$$

$$\alpha(nir) = \begin{bmatrix} 0.5; & T \leq 263.15 \\ 0.5 - 0.028(T - 263.15); & 263.15 < T < 273.15 \\ 0.22; & T \geq 273.15 \end{bmatrix} \quad \text{(Eqn. 13)}$$

PlaHab also models water ice coverage at a grid point ($f_{ice}$) to temperature based on empirical data (Thompson and Barron 1981). We had found the following fit (e.g. Ramirez 2020ab)(Equations 14 and 15):

$$f_{ice} = \begin{bmatrix} 1; & T \leq 239 \\ 1 - exp((T - 273.15)/12.5); & 239 < T < 273.15 \\ 0; & T \geq 273.15 \end{bmatrix} \quad \text{(Eqn. 14)}$$

Surface albedo at each grid point is calculated via the following (Equation 15):

$$a_s = (1 - f_c)\{(1 - f_o)a_l + f_o(f_i a_i + (1 - fi)a_o)\} + f_c a_c \quad \text{(Eqn. 15)}$$

Here, $a_s$, $a_c$, $a_o$, $a_i$, and $a_l$ are the surface, cloud, ocean, ice, and land albedo, respectively. Likewise, $f_c$, $f_o$, and $f_i$ are the cloud, ocean, and ice fraction, respectively. Following Fairén et al. (2012), at sub-freezing temperatures, the maximum value between ice and cloud albedo is chosen to prevent clouds from artificially darkening a bright ice-covered surface.

The vertical dimension is virtually absent (save for top-of-atmosphere quantities) in this model and so elevation variations of key quantities (i.e. surface temperature, precipitation, outgoing longwave radiation and TOA albedo) for our early Mars topography case were parameterized. Surface temperatures at different elevations (at a grid point) were calculated using 3D results for the lapse rate variation with height (i.e. pressure) on early Mars (Forget et al. 2013). Precipitation at higher altitudes were then computed by assuming a 7% decrease in precipitation rate per 1 degree Kelvin decrease (Neelin et al. 2022). I then fitted a formula for the top-of-atmosphere albedo and outgoing longwave radiation as a function of temperature as computed by PlaHab (see Appendix for more details on this parameterization).

PlaHab is extremely versatile in its input selection(Dietrich et al. 2023), enabling the simulation of $N_2$-$CO_2$-$H_2O$-$H_2$ atmospheres. The $N_2$, $CO_2$, and $H_2$ pressures can be varied whereas $H_2O$ pressures are self-consistently computed via the Clausius-Clapeyron equation (as explained above). Planetary, stellar and orbital parameters can also be varied, which includes the semi-major axis, rotation rate, planetary gravity, surface albedo, land/ocean coverage, obliquity, eccentricity, stellar type and stellar mass.

The model implements an explicit forward marching numerical scheme. For each time-step, spanning a few hours (default 6) of a planet's evolution, the new average surface temperature is updated for every grid point following equation (1), along with the resulting atmospheric and surface $H_2O$ and $CO_2$ inventories. For rapidly-rotating planets, convergence is achieved once average annual surface temperature variations decrease below ∼0.1 K and both $CO_2$ condensation/sublimation rates are balanced. However, I find that this method is insufficient for synchronously-rotating planets, producing extremely large (> ~10%) errors in top-of-atmosphere (TOA) energy balance. For such worlds, PlaHab achieves convergence when the net incoming flux equals that of the net outgoing thermal emission (at TOA), which produces almost no error (under 1%). The common inputs used for all simulations are summarized in Table 1.

Table 1: Simulation Inputs

| Input | Value | Units |
|---|---|---|
| $L$ | $5.9 \times 10^5$ | J/kg |
| $C_{po}$ | 1000 | g/kg/K |
| $P_o$ | 1 | bar |
| $\Omega_o$ | $7.27 \times 10^{-5}$ | rads/second |
| $m_o$ | 29 | g/mol |
| $FH_2O_o$ | 0.0155 | - |
| $C_d$ | $1.5 \times 10^{-3}$ | - |
| $v$ (near-surface) | 10 | m/s |
| $\Delta T$ (near-surface) | $7 \times 10^{-3}$ | K |
| $RH$ (surface) | 100% (aquaplanet), 0% dry planet, 77% for Mars and Earth | - |

## 3. RESULTS

### 3.1 Rapidly-Rotating Planets

First, I generate annually-averaged results for the present Earth to illustrate how well PlaHab simulates the current climate, approximated as 1 bar $N_2$ and 330 ppm $CO_2$ (Figure 1), a commonly-used proxy "modern Earth" composition (e.g. Kasting et al.1993).

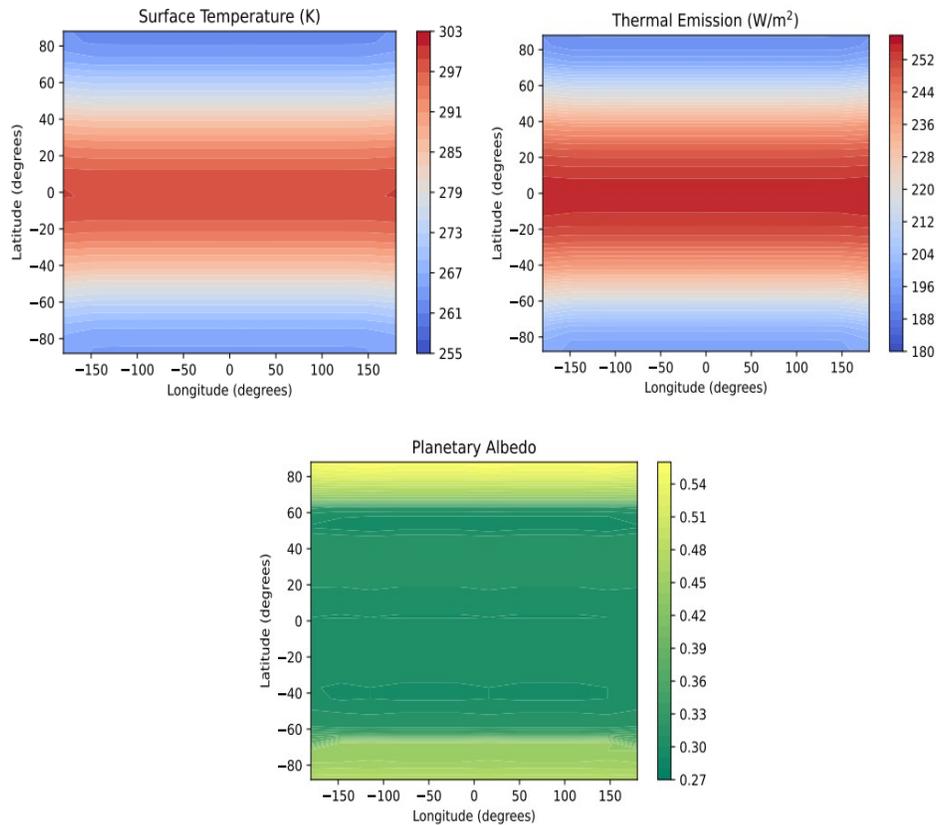

**Figure 1:** Plots of surface temperature (top left), thermal emission (top right) and planetary albedo (bottom) for "present" Earth ca. 1980 (330 ppm $CO_2$). The computed mean surface temperature, outgoing longwave radiation and planetary albedo are ~288 K, 238 W/m$^2$ and 0.3, respectively.

A flat topography was assumed, along with a 24 hour day and a 23.5 degree tilt. An Earth-like continental fraction as a function of latitude was assumed. Tropical surface temperatures exceed 300 K, producing the highest levels of thermal emission (> ~ 250 W/m$^2$. Model results exhibit no lateral heterogeneity for this flat topography case. In spite of greater cloud coverage at the warmer equator (Equation 5), the planetary albedo is highest near the poles (> 0.4) because of the presence of surface ice (Figure 1). Overall, the surface temperature distribution, planetary albedo, and thermal emission are similar to those obtained by observations and previous 1D EBM results (e.g. Trenberth et al. 2009; Stephens et al. 2015; Ramirez and Levi 2018).

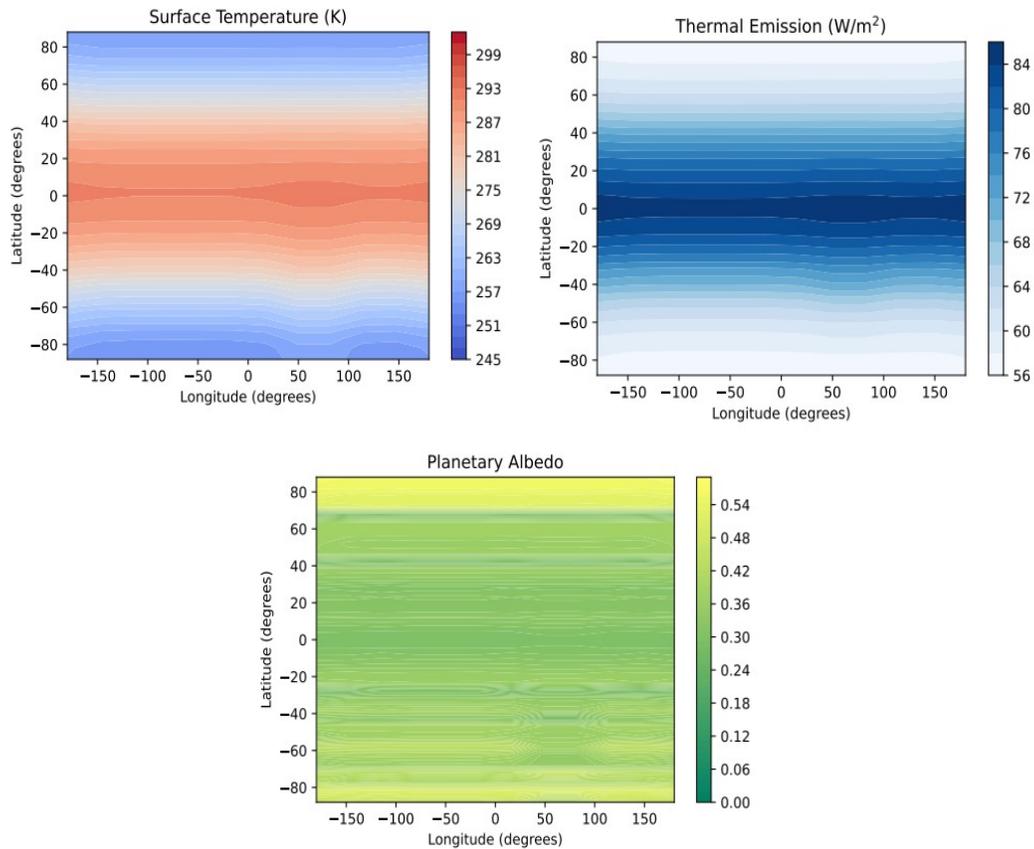

**Figure 2:** A simulated 2-bar 95% $CO_2$/5% $H_2$ early Mars (75% present solar flux) atmosphere at 0 degree obliquity with surface temperatures (upper left), thermal emission (upper right) and planetary albedo (bottom) for a case containing an ocean spanning 20N – 90N and a sea in Hellas Basin (center:~42S, 70E). Global mean annual surface temperature, thermal emission and planetary albedo are ~281 K, 74 W/m², and 0.33. Flat topography is assumed.

The next cases are for a warm (2 bar) early Martian atmosphere at 3.8 Ga (Figures 2 – 3). A zero degree obliquity and standard Martian day length (~24.6 Earth days) are assumed. The first set of figures represents a flat Mars (with no topography) with a large northern hemispheric ocean and a sea in Hellas Basin (Figure 2). This scenario is the 2D version of the same test case outlined in Ramirez et al. (2020), using the 1D EBM. Unlike the modern Earth simulation (Figure 1), these model results exhibit slight heterogeneity in the outgoing fluxes and surface temperature in a small region between the northern ocean edge and Hellas Basin (~0N, 75E), owing to a slight increase in regional atmospheric water vapor as moisture from both seas converge. Temperatures are warmest in the northern hemisphere (approaching 300 K) and subtropical regions as those locations are coincident with the ocean location and areas that receive significant moisture from the water sources.

In contrast, temperatures near the poles decrease to below 250 K (Figure 2). However, in this case, the northern ocean is partially frozen between ~65 – 90 N, producing locally cooler temperatures there (~265 K) and fewer clouds, resulting in lower albedo outgoing emission values (~60 W/m$^2$) and a higher planetary albedo (> 0.5). Thermal emission is lowest at both poles (Figure 2). Overall temperatures, fluxes, and albedo values (including globally mean-averaged outputs) are consistent with those from both Ramirez et al. (2020) as well as from 1D radiative-convective climate modeling simulations (Ramirez 2017; Wordsworth et al. 2017).

In the next set of plots (Figure 3), I showcase the impact of topography in PlaHab and model the same atmosphere for a dry planet (no water clouds, ocean nor sea). I've included rough topographical representations for the Tharsis bulge and Hellas basin (e.g. Bernhard et al. 2023). I've also incorporated the average elevations of the southern highlands (~ the southern 60% of the planet) and northern lowlands, which are ~ 2km above the mean and 3 km below it (e.g. Smith et al. 1999; Zuber 2000). A rough sketch of the assumed topography is given in the Appendix.

Topography still impacts regional surface temperatures even in the absence of a rain shadow effect (Figure 3). The highest temperatures of the low-lying Hellas Basin rival those at higher latitudes, along with a somewhat higher ("leakier") outgoing longwave radiation. Likewise, the Tharsis region exhibits temperatures of ~280 K, as compared to surrounding regions with temperatures at least 10 K higher. Relative to Figure 2, the albedo is relatively uniform across the planet, owing to the lack of surface ice, water and clouds. In addition, mean surface temperatures between the moist (Figure 2) and dry (Figure 3) scenarios are very similar (to within ~2 – 3 K), which is the same result obtained by recent GCM simulations (Guzewich et al. 2021).

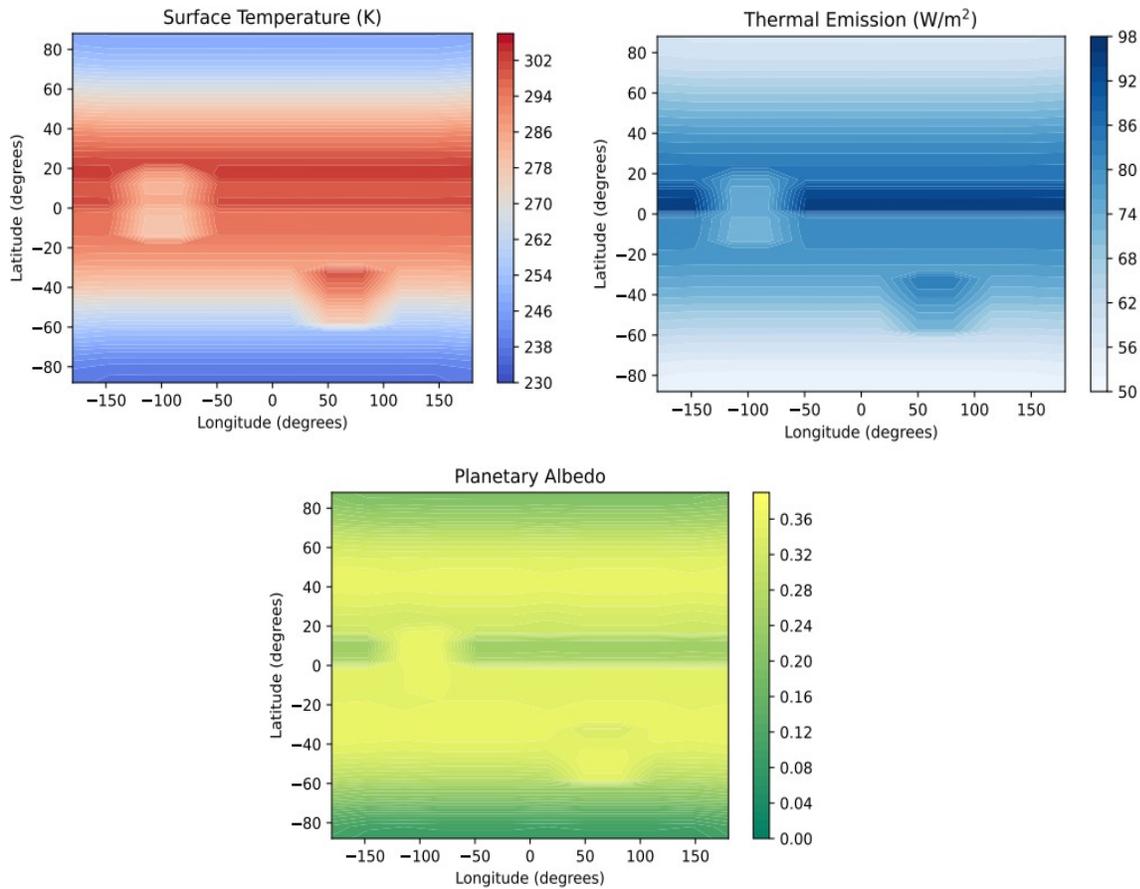

**Figure 3:** A simulated 2-bar 95% $CO_2$/5% $H_2$ early Mars (75% present solar flux) atmosphere at 0 degree obliquity with surface temperatures (upper left), thermal emission (upper right) and planetary albedo (bottom) for a dry case with no water cycle and with Hellas Crater and Tharsis bulge (center: -1S, 100W) topography. Global mean annual surface temperature, thermal emission and planetary albedo are ~284 K, 78 W/m², and 0.30.

### 3.2 Synchronously-Rotating Planets

The next several sets of plots (Figures 4 – 7) represent PlaHab Trappist-1 Habitable Atmosphere Intercomparison (THAI) project results for the two Benchmark (Figures 4 and 5) and two Habitable Cases (Figures 6 and 7), respectively, for TRAPPIST-1e (incident flux of 900 W/m² from a 2600 K M-star) (Fauchez et al. 2020) with assumed values for the orbital period (6.1 days), planetary mass (0.772 Earth's) and gravitational acceleration (9.12 m/s²). I also compare against recent GCM and EBM simulations for a synchronously-rotating Proxima Centauri b (956 W/m² of incident flux from a 3000 K M-star), assuming an orbital period of 11.5 days, gravitational acceleration of 10.9 m/s², and a mass of 1.4 Earths(Figures 8 and 9; Turbet et al. 2016; Okuya et al. 2019). Flat topography and zero obliquity are assumed for all of these cases. As predicted for synchronously-rotating planets, a distinct "eyeball" pattern of increased instellation forms on the dayside areas surrounding the substellar point (e.g.

Pierrehumbert 2010; Yang et al. 2013; 2014) (Figures 4 - 9). A fraction of this absorbed energy is transported to the nightside. In all these cases, the substellar point corresponds to a longitude and latitude of zero degrees.

### 3.2.1 THAI Benchmark cases

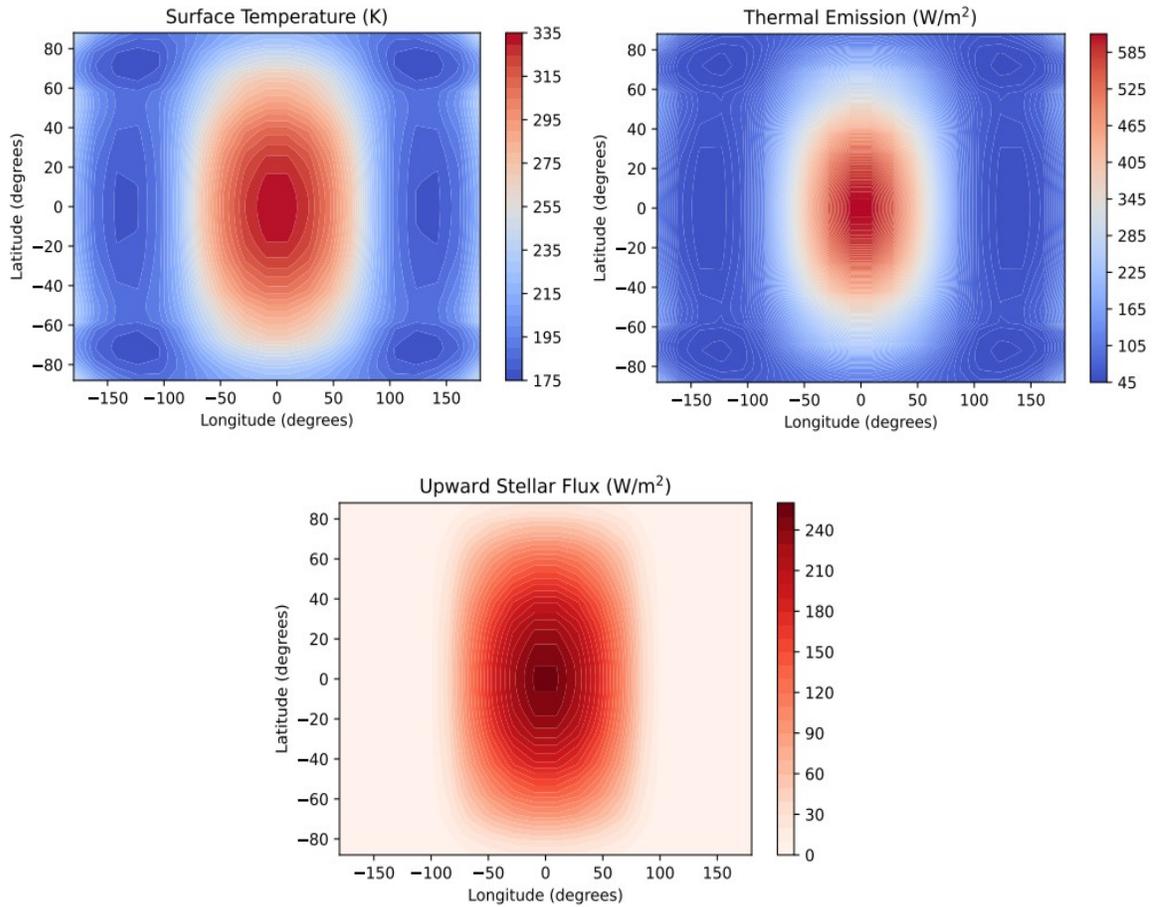

**Figure 4:** TRAPPIST-1e 1-bar $N_2$ 400 ppm $CO_2$ atmosphere dry planet (THAI Benchmark 1) with plots of surface temperature (upper left), thermal emission (upper right) and upward stellar flux (bottom). Global mean surface temperature, thermal emission and planetary albedo are 235 K, 193 W/m$^2$, 0.143, respectively.

Benchmarks 1 and 2 are dry land planets (no water vapor/ice, nor associated clouds)with an assumed surface albedo of 0.3 (Fauchez et al. 2020). In the 1 bar $N_2$ 400 ppm Benchmark 1 case (Figure 4), daytime surface temperatures peak to 335 K at the substellar point whereas the lowest are ~ 175 K on the nightside (mean surface temperature of 235 K). Thermal emission peaks and lows are ~612 and 40 W/m$^2$, respectively, whereas the planetary albedo peaks at ~0.287 (Figure 4). The unusually high thermal emission owes to both the low greenhouse gas concentrations and the lack of atmospheric water vapor absorption. The mean atmospheric temperature (235 K) is slightly higher than that predicted by recent GCMs (average of ~217 K; Turbet et al. 2022) or by a recent tidally-locked 1D EBM (T = 224 K; Haqq-Misra et al. 2023). However, the slightly higher mean surface temperature computed by PlaHab is still consistent with expectations as this temperature is still significantly cooler than the Earth (by ~53 K) given a *net* incident flux ~81% that of our own planet.

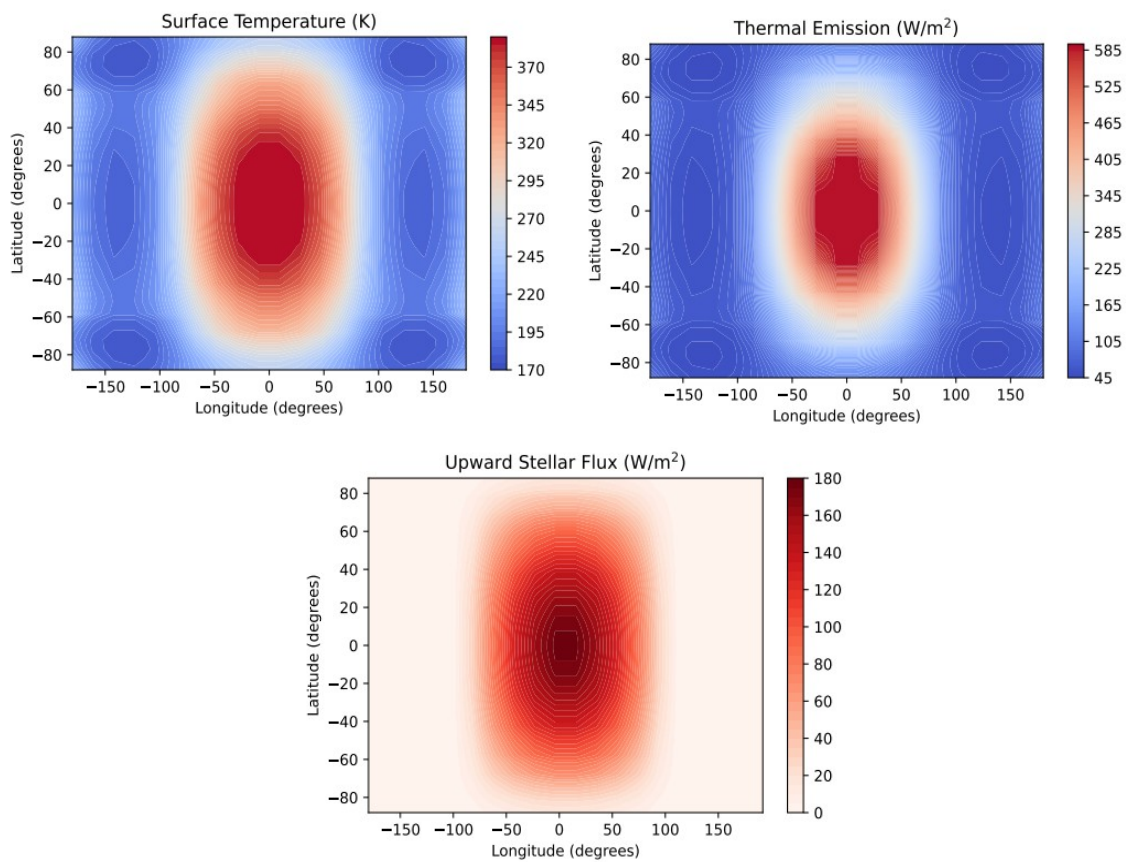

**Figure 5:** TRAPPIST-1e 1-bar $CO_2$ atmosphere dry planet (THAI Benchmark 2) with plots of surface temperature (upper left), thermal emission (upper right) and upward stellar flux (bottom). Global mean surface temperature, thermal emission and planetary albedo are 262 K, 202 W/m$^2$, 0.102, respectively.

Temperatures are even higher in the 1 bar $CO_2$ case (Figure 5). With maximum (minimum) temperatures of 390 K (175 K), maximum (minimum) thermal emission of 589 W/m$^2$ (48 W/m$^2$), and maximum planetary albedo of ~ 0.4 (in a thin sliver at the poles). The mean atmospheric temperature is 262 K, which is somewhat higher than the ~239 – 245 computed by recent GCMs (~242 K average; Turbet et al. 2022)and much lower than the ~286 K computed by Haqq-Misra et al. (2022). As compared to the dry 1-bar $N_2$ case (Figure 4), there is a decrease in the planetary albedo (reduced reflected stellar flux) corresponding to the increase in absorption (Figure 5).

**3.2.2 THAI Waterworld cases**

In contrast to the Benchmark cases, Habitable Cases are aquaplanets with assumed albedo values of 0.06 for ocean and 0.25 for ice, with an effective heat capacity of 4x10$^7$ Jm$^{-2}$K$^{-1}$ (Fauchez et al. 2020). The same atmospheric compositions assumed for the Benchmarks are also implemented in the Habitable Cases (Fauchez et al. 2020) (Figure 6).

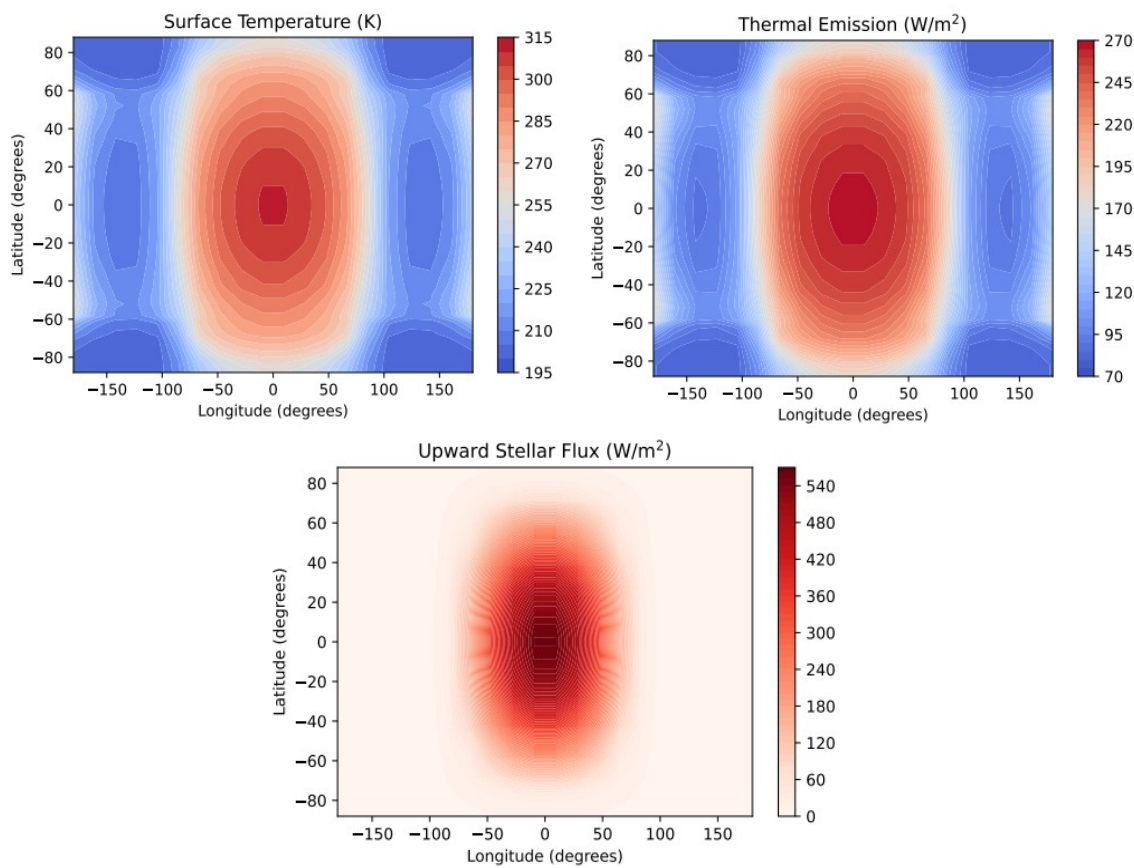

**Figure 6:** TRAPPIST-1e 1 bar $N_2$ and 400 ppm atmosphere aqua planet (THAI Habitable Case 1) with plots of surface temperature (upper left), thermal emission (upper right) and upward stellar flux (bottom). Global mean surface temperature, thermal emission and planetary albedo are 250 K, 169 W/m$^2$ and 0.245, respectively.

Surface temperatures for Habitable Case 1 range from ~310 K (at the substellar point) to 201 K whereas the corresponding thermal emission spans from ~81 to 267 W/m$^2$ (Figure 6). The maximum planetary albedo is ~0.63, again at the substellar point. An increase in dayside-to-nightside transport is observed in this case relative to the dry equivalent (Figure 4), as significantly higher temperatures (10s of K) characterize the nightside in this moist case (Figure 4; Figure 6). The upward stellar fluxes are much higher higher here than in the dry case (Figure 4) given the build up of highly-reflective dayside clouds near and around the substellar point(Yang et al. 2013; 2014). This presence of water vapor (~1% at the surface) increases the planetary albedo, but this is offset by enhanced absorption, resulting in an overall increase in mean surface temperature (250 K vs. 235 K). PlaHab's calculated mean surface temperature is similar to the high end estimate from a resent ensemble of GCMs (~ 232 – 246; average: 241 K ; Sergeev et al. 2022)(Figure 6). Both sets of results are warmer than the 226 K computed by the 1D EBM of Haqq-Misra et al. (2022).

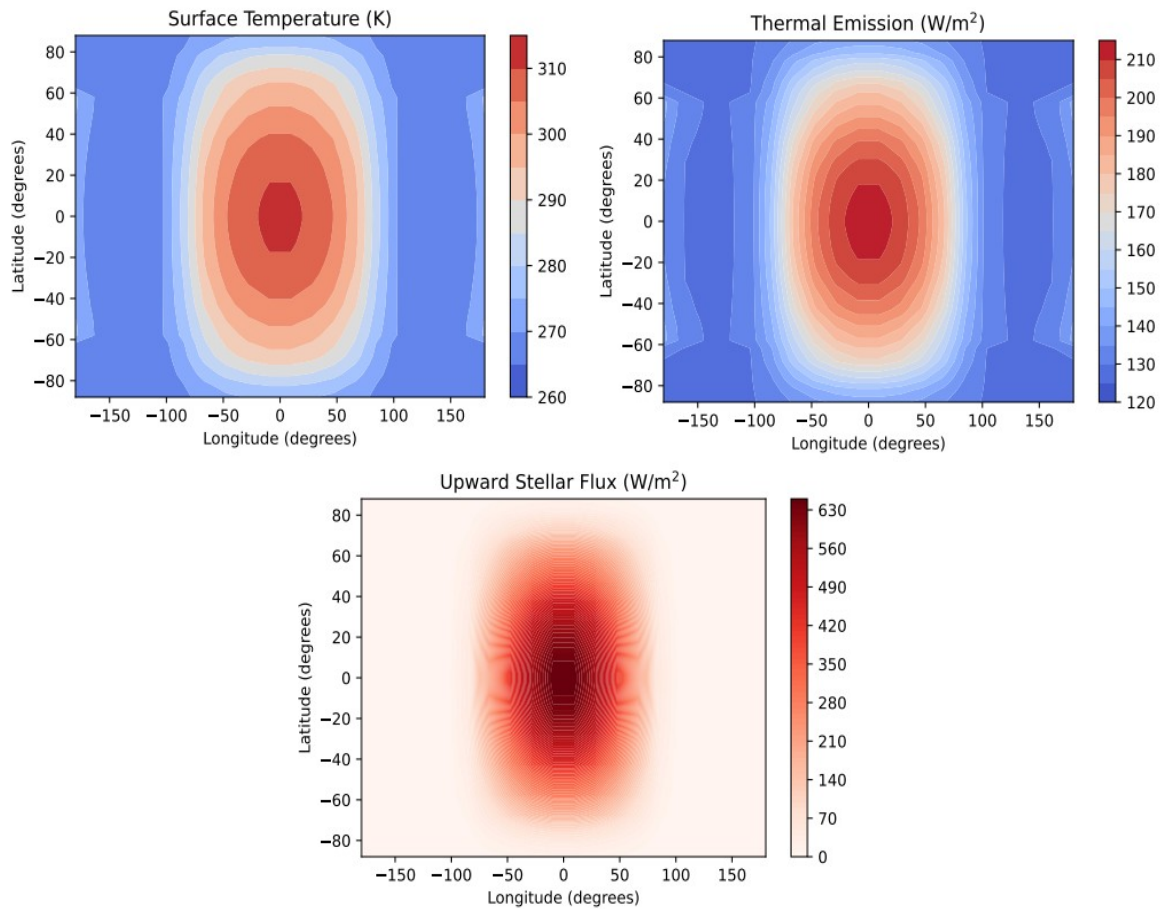

**Figure 7:** TRAPPIST-1e 1 bar $CO_2$ atmosphere aqua planet (THAI Habitable Case 2) with plots of surface temperature (upper left), thermal emission (upper right) and upward stellar flux (bottom). Global mean surface temperature, thermal emission and planetary albedo are ~282 K, 155 W/m$^2$ and 0.312 respectively

In comparison, Habitable Case 2 is the hottest scenario in this THAI intercomparison (Figure 7). Temperatures decrease from ~311 K on the dayside substellar point to ~267 K on the nightside. The thermal emission is lowest on the nightside at ~129 W/m² and highest on the dayside at 212 W/m² (Figure 7). The planetary albedo peaks at ~ 0.73 with an upward stellar flux reaching ~650 W/m², again given the efficacy of dayside clouds. (Figure 7). This is also the wettest THAI atmosphere, as the surface water vapor volume mixing ratio exceeds ~1.8%. This enhanced absorption is manifested by relatively low planetary albedo values and a larger dayside "eyeball" with even more pronounced thermal connectivity between the day- and nightsides (Figure 7). PlaHab's calculated mean surface temperature is ~282 K, which is comparable to both the ~283 K GCM average (ranging from ~ 271 – 295 K) of Sergeev et al. (2022) and somewhat lower than the ~287 K computed by Haqq-Misra et al. (2022).

### 3.2.3 Proxima Centauri b

I also show a similar comparison against Proxima Centauri b, assuming aquaplanets with 1 bar $N_2$ 376 ppm $CO_2$ and 1 bar $CO_2$ atmospheres (Turbet et sl. 2016). For the 376 ppm $CO_2$ atmosphere, the highest dayside temperature as computed by PlaHab is ~ 307 K, only slightly higher than the maximum dayside temperature of ~303 K in Turbet et al. (2016). In comparison, both of these estimates are slightly lower than the 315 K substellar point temperature obtained by the 2D EBM of Okula et al. (2019). The lowest nightside temperature I compute is ~ 203 K (Figure 8). The thermal emission ranges from ~87 W/m² to a maximum of ~264 W/m² at the substellar point. The planetary albedo also reaches a maximum value of ~0.64 near the substellar point, with upward stellar fluxes exceeding ~ 600 W/m² (Figure 8). Overall, the temperatures and trends are comparable to those of the GCM of Turbet et al. (2016). For example, PlaHab computes a global mean surface temperature of ~254 K (Figure 8), which is very similar to the 252 K value for Turbet et al. (2016). This is ~34 K lower than our planet's mean surface temperature for a net incident flux ~75% that of Earth's.

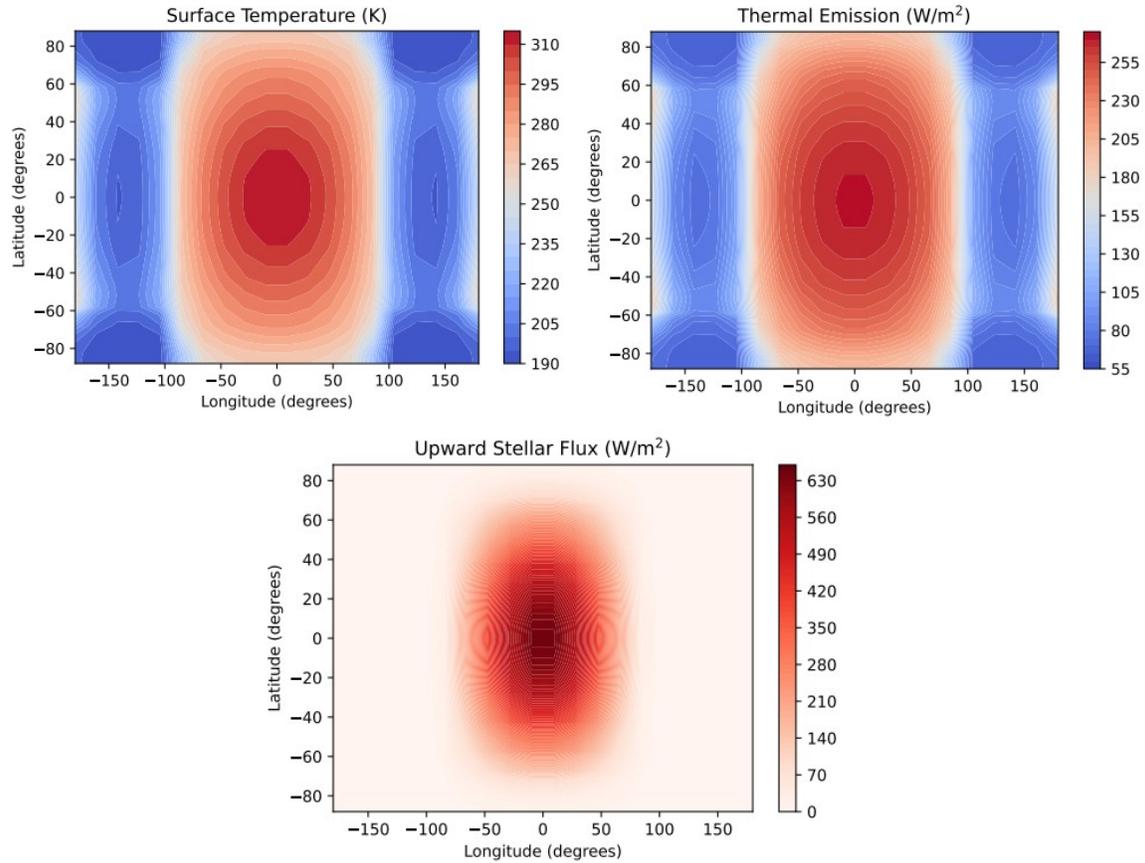

**Figure 8:** Proxima Centauri b 1 bar $N_2$ and 376 ppm $CO_2$ atmosphere aqua planet with plots of surface temperature (upper left), thermal emission (upper right) and upward stellar flux (bottom). Global mean surface temperature, thermal emission and planetary albedo are 254 K, 176 W/m$^2$, and 0.260, respectively.

Our last comparison is for the 1 bar $CO_2$ atmosphere of Turbet et al. (2016). This is the warmest atmosphere in this paper with the highest degree of transport, resulting in greatly diminished meridional and zonal temperature gradients. Substellar point maximum temperatures reach ~ 301 K with a nightside minimum that is only ~ 3 K lower (Figure 9). Thermal emission ranges from 154 – 158 W/m$^2$ whereas the planetary albedo maximum is ~ 0.71 and the upward stellar flux maximizes at ~646 W/m$^2$ near the substellar point (Figure 9). The average mean surface temperature is 299 K (Figure 9), nearly identical to the 300 K value in Turbet et al. (2019). The latter study also found greatly reduced temperature gradients, with zonally-averaged equator-pole temperatures varying by less than 10 degrees (Figure 10). Again, these numbers and trends are consistent with the equivalent case in Turbet et al. (2016), even if our gradients are slightly smaller than that in the Turbet et al. study.

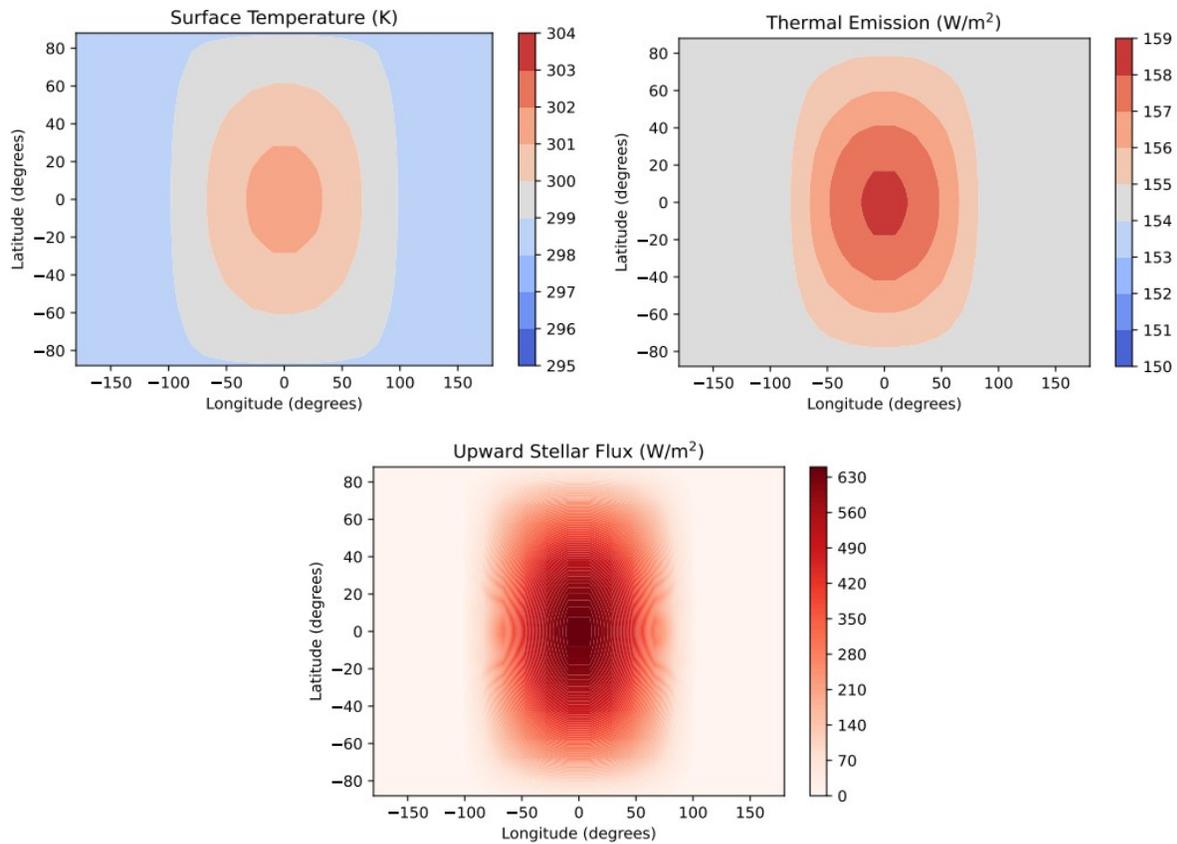

**Figure 9:** Proxima Centauri b 1 bar $CO_2$ atmosphere aqua planet with plots of surface temperature (upper left), thermal emission (upper right) and upward stellar flux (bottom). Global mean surface temperature, thermal emission and planetary albedo are 299 K, 155 W/m² and 0.344, respectively.

A summary of all the main comparison results is given in Table 2. Like the recent GCM simulations (Sergeev et al. 2022; Turbet et al. 2022), top-of-atmosphere energy balance (net incident flux = net outgoing infrared radiation) is achieved in all cases (Table 2), indicating the robustness of the convergence algorithm.

**Table 2: Globally-Averaged Mean Surface Temperature, Outgoing Longwave Radiation, and Planetary Albedo Intercomparison Summary**

| Case | Composition | T (K) | Thermal emission TOA (W/m$^2$) | Planetary Albedo |
|---|---|---|---|---|
| Earth | 1 bar $N_2$, 330 ppm $CO_2$ | 288 | 238 | 0.30 |
| Mars 3.8 Ga(flat) | 2 bar 95% $CO_2$, 5% $H_2$ | 281 | 74 | 0.33 |
| Mars 3.8 Ga | 2 bar 95% $CO_2$, 5% $H_2$ | 284 | 78 | 0.30 |
| Benchmark 1 | 1 bar $N_2$, 400 ppm $CO_2$ | 235 | 193 | 0.143 |
| Benchmark 2 | 1 bar $CO_2$ | 262 | 202 | 0.102 |
| Habitable Case 1 | 1 bar $N_2$, 400 ppm $CO_2$ | 250 | 169 | 0.245 |
| Habitable Case 2 | 1 bar $CO_2$ | 282 | 155 | 0.312 |
| Proxima Centauri b 1 | 1 bar $N_2$ 376 ppm $CO_2$ | 254 | 176 | 0.260 |
| Proxima Centauri b 2 | 1 bar $CO_2$ | 299 | 155 | 0.344 |

## 4. DISCUSSION

### 4.1 Model Summary

As shown above, PlaHab successfully simulates both rotating and synchronously-rotating atmospheres, yielding answers that are largely consistent with other models, including GCMs, to within ~0 - 15 K for mean surface temperatures (~0 – 5 K for the Habitable Cases) using self-consistently computed parameterizations that require no further tuning. For comparison, mean surface temperature discrepancies between a recent 1D tidally-locked EBM and calculated GCMs approached ~40 K for Benchmark 2(Haqq-Misra and Hawyworth 2022). As also shown by Haqq-Misra and Hayworth (2022), standard 1D EBMs that are not properly adapted to simulate synchronously-rotating planets generate even larger errors.  PlaHab also reproduces the proper values for mean surface temperature, planetary albedo, and outgoing thermal flux for present Earth. Therefore, equations 1 - 3 can adequately model the heat transport for most terrestrial planets (whether synchronously-rotating or not).

**4.2 The importance (or lack thereof) of $CO_2$ condensation**

These simulations offer additional insights. Whereas the THAI protocol maintains consistency by disabling $CO_2$ condensation (Fauchez et al. 2020), I have included such effects (as explained in Methods) to assess their importance for these comparisons. I find that $CO_2$ does not condense for the vast majority of these simulations, save for the 1 bar $CO_2$ Benchmark 2 case, which produced a non-negligible amount, with ~10% $CO_2$ condensing on to the nightside poles to combine with a global $CO_2$ cloud fraction of ~29%. Sensitivity studies revealed that this negligibly affected global mean surface temperatures (< 1 K), although this result should be revisited with other models, including GCMs, with proper dynamics. In contrast, water vapor absorption was sufficiently intense to prevent surface $CO_2$ condensation in the 1-bar $CO_2$ Habitable 2 case. Likewise, the 1-bar $CO_2$ Proxima Centauri b and early Mars cases were simply too hot to trigger $CO_2$ condensation. In the case of early Mars, the additional warming from $CO_2$-$H_2$ collision-induced absorption helped prevent atmospheric collapse (e.g. Ramirez et al. 2014; Ramirez 2017; Wordsworth et al. 2017) Thus, PlaHab predicts that most of these comparisons would not have needed to include $CO_2$ condensation, save for the Benchmark 2 scenario (which should be re-evaluated by future GCMs that incorporate proper $CO_2$ condensation physics).

**4.3 Comparing Planetary Albedo as computed by PlaHab and GCMs**

The PlaHab THAI and Proxima Centauri b simulations reveal that the planetary albedo is higher in the $CO_2$ rich (1 bar) cases over the $CO_2$-poor (400 ppm) ones (Table 2). This makes sense on first glance, not only because of the increase in dayside cloud coverage at higher temperatures, but because of the higher scattering cross-section of $CO_2$ over terrestrial air (e.g. Kasting et al. 1993). However, this is not necessarily what the GCMs find. Although the PlaHab-calculated planetary albedo for Habitable Case 1 (~0.245) is similar to that obtained by GCMs (Table 2; ~0.22 – 0.28; e.g. Sergeev et al. 2022), PlaHab and GCM-computed planetary albedoes (and therefore, outgoing longwave radiation) diverge somewhat for the remaining THAI cases. For instance, PlaHab computes a higher albedo (~0.312) for Habitable Case 2 than that computed by recent GCMs (~0.15 – 0.26)(Sergeev et al. 2022). Thus, for a given atmospheric composition, energy is more efficiently absorbed in the EBM relative to GCMs for Habitable Case 2. I attribute this discrepancy to the common GCM prediction of decreasing low level water cloud opacity at sufficiently high temperatures, resulting in a concomitant decrease in the planetary albedo (e.g. Brient and Bony 2012). This effect is not modeled in PlaHab, which assumes that tropospheric cloud fractions only increase as temperatures do. This is why the PlaHab planetary albedo increases from Habitable Case 1 to 2 (~0.26 to 0.31). In contrast, the GCMs produce the opposite trend, with the planetary albedo decreasing somewhat from Habitable Case 1 to Habitable Case 2 (Turbet et al. 2022; Sergeev et al. 2022).

Interestingly, the opposite phenomenon occurs in the cloudless dry Benchmark cases. That is, planetary albedo as computed by PlaHab *decreases* (0.143 to 0.102; Table 2) from the low to high $CO_2$ Benchmark Case (instead of increasing). In this case, however, GCMs *also* find the same decreasing trend (Turbet et al. 2022) that PlaHab does, suggesting again that the different trends observed above for the Habitable Cases are apparently related to how clouds are modeled. That being said, this time the GCMs obtain planetary albedo values that are somewhat *higher* (~ 0.2 – 0.3) than PlaHab's. So, what is causing the higher albedo in the GCM benchmark cases given the absence of clouds in both PlaHab and the GCMs for these latter scenarios(assuming all else equal, including radiative transfer)? It is possible that assuming a heat transport over dry land that is half as effective as that over oceans, as implemented in both Pollard et al. (1983) and Okuya et al. (2019), is incorrect, incomplete or should not be applied on a planetary scale. The possibility of incorrect diffusive transport assumptions for dry planets may also explain why agreement between PlaHab and GCMs is greater for water-rich planets(mean surface temperatures agree to within 5 K) over the dry Benchmark cases (up to a ~17 K discrepancy for Benchmark 2). Future work should revisit and test the validity of previous transport assumptions in the context of dry synchronously-rotating planets.

**4.4 Comparison of Spatial Temperature Gradients with GCMs and Observations**

Ultimately, the validity of a 2D EBM depends on its ability to faithfully reproduce mean-averaged *and* spatially distributed values that are consistent with observations or what a GCM may produce. For this purpose, I have compared PlaHab zonally-averaged surface temperatures for many of the cases in this paper against either GCMs or observations (Figure 10). For the Habitable and Benchmark cases, I had decided to compare against the ROCKE-3D GCM (Way et al. 2017), as it consistently produced surface temperature results that were in the middle of those computed by the THAI model ensemble (Sergeev et al. 2022; Turbet et al. 2022). For the Proxima Centauri b comparison, I had compared PlaHab against the LMD GCM used in the original calculation (Turbet et al. 2016).

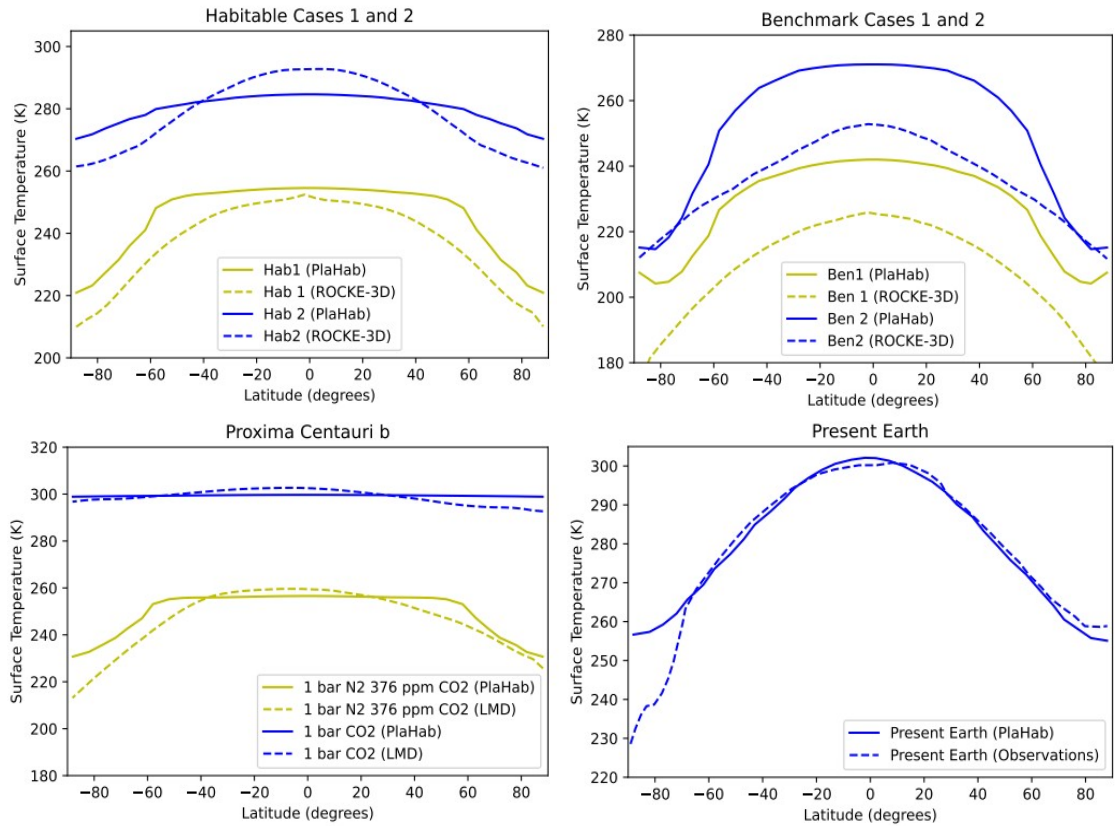

**Figure 10:** Zonally-averaged surface temperature comparison of PlaHab for (upper left) both Habitable and (upper right) Benchmark Cases against ROCKE-3D, (lower left) both Proxima Centauri b cases against LMD and versus (lower right) present Earth observations. Habitable Cases 1 and 2 contain 1bar $N_2$ with 400 ppm $CO_2$ and 1 bar $CO_2$, respectively. Benchmark Cases are similar to Habitable Cases minus the water vapor. The present Earth PlaHab simulation contained 1 bar $N_2$ and 330 ppm $CO_2$.

Again, agreement is significantly better for the wet planets (i.e. Habitable Cases, Present Earth, Proxima Centauri b) than for the Benchmark Cases. Not only are mean surface temperatures similar for all wet planets (as discussed in Results), but the temperature gradient trends are mostly consistent, with modest differences. The present Earth fit is at least as good (if not slightly better) in PlaHab than for our equivalent 1D EBM calculation (Ramirez and Levi 2018). As discussed in Ramirez and Levi (2018), the discrepancies near the south pole are largely attributed to the high topography on the real Earth, which would produce lower surface temperatures than those predicted by my simulated flat surface cases.

Habitable Case 2 exhibits an equator-pole temperature gradient discrepancy of ~15 K between PlaHab and ROCKE-3D, with similar mean surface temperatures (to within ~ 1 K). For comparison,

Habitable Case 1 equator-pole temperature gradients are ~ 42 K in ROCKE-3D and 34 K in PlaHab (~8 K difference). The Proxima Centauri b LMD simulations exhibited significant northern-southern hemispheric asymmetry (Figure 10) in spite of the flat topography and zero obliquity assumptions, making a direct comparison more difficult. However the *mean* equator-pole temperature gradient for the 376 ppm CO2 Proxima Centauri b case are ~41 and 25 K (~16 K difference, decreasing to ~3K at the north pole), respectively, for LMD and PlaHab (Figure 10).

Overall, the level of difference (~10 – 60%)in the equator-pole gradients for our wet cases is consistent with those that may be found across GCMs (e.g. see Figure 12 in Komacek and Abbot 2019). As a further demonstration, I have computed the equator-to-pole surface temperature gradients for a planet with a "present Earth" proxy atmospheric composition (i.e. 1 bar N2, 330 ppm $CO_2$)for different rotation rates and compared against the same calculation made by an idealized GCM (Kaspi and Showman 2015)(Figure 11). The agreement is quite good, showcasing that PlaHab operates reasonably well over a large range of rotation rates (or at the least is consistent with other complex models). PlaHab also reproduces the "kink" at ~20 – 30% of Earth's rotation rate that 3-D GCMs also obtain (Kaspi and Showman 2015; Komacek and Abbot 2019). This discontinuity is attributed to the high dependence between rotation rate and transport at high rotation rates, which contrasts with the relative lack of sensitivity in these quantities at low rotation rates (Kaspi and Showman 2015).

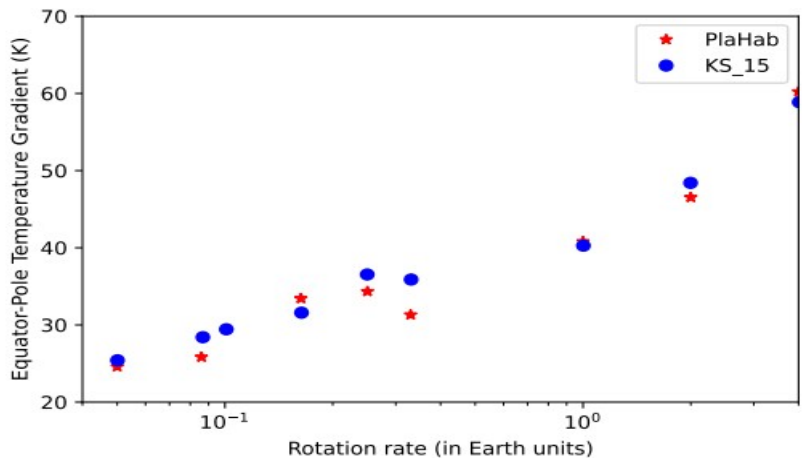

**Figure 11:** Equator-pole temperature gradient as a function of rotation rate for a planet with a modern Earth atmospheric composition comparing (red stars) PlaHab and (blue dots) Kaspi and Showman (2015) data points.

**CONCLUSION**

A tool, properly implemented within its limitations, is very useful. A 2D EBM is no exception, approximating more of the complexity of a full-GCM, but at a reduced complexity and computational cost. Building upon previous work (Okuya et al. 2019), I have developed a 2D EBM (PlaHab) that can simulate the atmospheres of both rapidly-rotating and synchronously-rotating terrestrial planets with $N_2$-$CO_2$-$H_2O$-$H_2$ atmospheres (including explicit $CO_2$ condensation) orbiting main-sequence stars (with and without topography). PlaHab's agreement with GCMs in calculated mean surface temperatures for rapidly-rotating (Earth, Mars) and water-rich synchronously rotating habitable planets (TRAPPIST-1e, Proxima Centauri) was within ~ 5 K. Slightly worse agreement (< ~15 K) with the dry Benchmark cases suggests that further work is needed to develop improved diffusive parameterizations for dry planets. That said, the errors in simulating dry planets is probably expected in an EBM that was specifically designed to simulate potentially habitable planets (those with at least *some* surface and/or atmospheric $H_2O$). Overall, the results here show a marked improvement over standard 1D EBMs (e.g. Haqq-Misra et al. 2016; Ramirez 2020ab), including tidally-locked ones (e.g. Haqq-Misra and Hayworth 2022), in achieving congruence with GCMs, especially for synchronously-rotating planets.

To conclude, this paper further demonstrates the usefulness of 2D EBMs for simulating both rapidly-rotating and synchronously-rotating planets and are a valuable addition to the hierarchy of planetary/exoplanetary climate models (i.e. RCMs, 1D EBMs, and GCMs). Their simplicity permits realistic answers to be computed with far fewer computational resources, enabling a greater exploration of parameter space, which in combination with other models, can enhance the workflow of planetary climate modelers.

**Acknowledgments**

The author thanks Dorian Abbot and an anonymous referee for excellent reviews that greatly improved the scientific clarity of the manuscript. The author also acknowledges enlightening discussions with Ayaka Okuya on energy balance in 2D energy balance climate models and with Jun Yang and Thaddeus Komacek on rotation rates in tidally-locked planets. The author also gives a big thanks to Thomas Fauchez, Thaddeus Komacek, Martin Turbet and Michael Way for sharing their GCM output, which helped validate the EBM. The author had fruitful discussions with the following colleagues of the FILLET/CUISINES EBM modeling intercomparison project: Russell Deitrick, Thomas Fauchez, Jacob Haqq-Misra, Shintaro Kadoya, Paolo Matteo Simonetti and Vidya Venkatesan.

THAI data have been obtained from https://ckan.emac.gsfc.nasa.gov/organization/thai, a data repository of the Sellers Exoplanet Environments Collaboration (SEEC), which is funded in part by the NASA Planetary Science Divisions Internal Scientist Funding Model.

PlaHab will be publicly available via the University of Central Florida's institutional repository, STARS (stars.library.ucf.edu), an open-access digital repository for the research and scholarly output of the University of Central Florida. This institutional repository operates on the bepress Digital Commons platform, a comprehensive hosted solution for storing, managing, and sharing data. The service offers unlimited storage, customizable metadata, authorization and access control tools, long-term stable URLs, on-demand metrics, and search engine indexing. The data and metadata are managed by the PI in consultation with repository staff and librarians. Data appropriate for public release will be openly published and archived through STARS. Those data requiring access restriction periods prior to open access publication through the repository will be appropriately embargoed and/or restricted based on user id, IP range, or domain.

# APPENDIX

# A New 2D Energy Balance Model For Simulating the Climates of Rapidly- and Slowly-Rotating Terrestrial Planets

## A1. Longitudinal and Latitudinal Parameterizations for Rapidly- and Slowly-Rotating Terrestrial Planets

As discussed in the main text, the latitudinal ($D_{o,1}$) and longitudinal ($D_{o,2}$) diffusion coefficients were fitted to the following quadratic equation (Eqn. A1) :

$$D_o,(_{1,2}) = a_1\eta^2 + a_2\eta + a_3 \qquad \text{(Eqn. A1)}$$

Here, $\eta = \Omega/\Omega_o$.

The following tables summarize the fitting parameters for rapidly- and slowly-rotating planets (Tables A1 – A2).

### Table A1: Equation A1 Parameters for Slowly-rotating Planets

| For $D_{o,1}$: | $0.05 < \eta < 0.0866$ | $0.0866 \leq \eta < 0.163$ | $0.163 \leq \eta < 0.33$ | |
|---|---|---|---|---|
| $a_1$ | 0 | 0 | -0.021164567416891 | |
| $a_2$ | 0.000136612021858 | 0.013219895287958 | 0.049775449101797 | |
| $a_3$ | 0.000328169398907 | -0.000804842931937 | -0.006121076811893 | |
| | | | | |
| For $D_{o,2}$: | $0.05 < \eta < 0.0866$ | $0.0866 \leq \eta < 0.163$ | $0.163 \leq \eta < 0.25$ | $0.25 \leq \eta < 0.33$ |
| $a_1$ | 0 | 0 | 0 | 0 |
| $a_2$ | -49.3819672131148 | -4.30130890052356 | -12.3356321839081 | -1.75 |
| $a_3$ | 5.68509836065574 | 1.78111335078534 | 3.78390804597701 | 1.1375 |

### Table A2: Equation A1 Parameters for Rapidly-rotating Planets

| For $D_{o,1}$: | $1 < \eta < 1.5$ | $1.5 \leq \eta < 3$ | $3 \leq \eta < 4$ |
|---|---|---|---|
| $a_1$ | 0 | 0 | 0 |
| $a_2$ | 1.1 | 1.2333333333 | 1.8 |
| $a_3$ | -0.6 | -0.8 | -2.5 |

For Earth (Table A2), eqn. A1 yields 0.5 for $D_{o,1}$, which is the same value assumed in the text.

For rapidly-rotating planets, the temperature distribution is largely independent of the longitudinal heat transport, yielding a very small $D_{o,2}$ for such worlds (e.g. Okuya et al. 2019). Thus, the near-zero $D_{o,2}$ values assumed for Earth mentioned in the main text were also implemented for Mars and other rapid-rotators.

**A2. Dayside Cloud Albedo Parameterization**

The dayside cloud albedo was parameterized as follows (Eqn. A2):

$$cpalb(\theta, \phi) = min\left(c_p log\left(\frac{F_C}{F_E} + 1\right), 1\right) \tag{Eqn. A2}$$

As mentioned in the main text, $c_p$ is the dayside cloud enhancement factor, which is a function of temperature (Eqn. A3):

$$c_p = max\left(0.587787827608533 ln(T_{ann}) - 2.95318972452402, 0.3\right) \tag{Eqn. A3}$$

Here, $T_{ann}$ is the annually-averaged mean surface temperature. Thus, the default value for $c_p$ is 0.3 at temperatures below ~250 K, increasing to 0.4 at 300 K.

**A3. Planetary Albedo and OLR parameterization for Mars Topography Simulation**

The baseline simulations had assumed a flat planet and both OLR and PALB had been self-consistently calculated. After assuming a lapse rate (discussed in the main text), the OLR and PALB at higher elevations were computed using the following model-derived parameterizations at every grid point (Eqns. A4 - A5):

$$OLR(\theta, \phi) = 0.4968591205 T - 65.377886522 \tag{Eqn. A4}$$

$$PALB(\theta, \phi) = -0.000116269879887 T^2 + 0.06626385910216 T - 9.085082598 \tag{Eqn. A5}$$

Here, T is the surface temperature at a given elevation.

## A4. Input Map for Mars Topography Simulation

The topography assumed for Figure 3 was overlain on this publicly-available MOLA map (Figure A1):

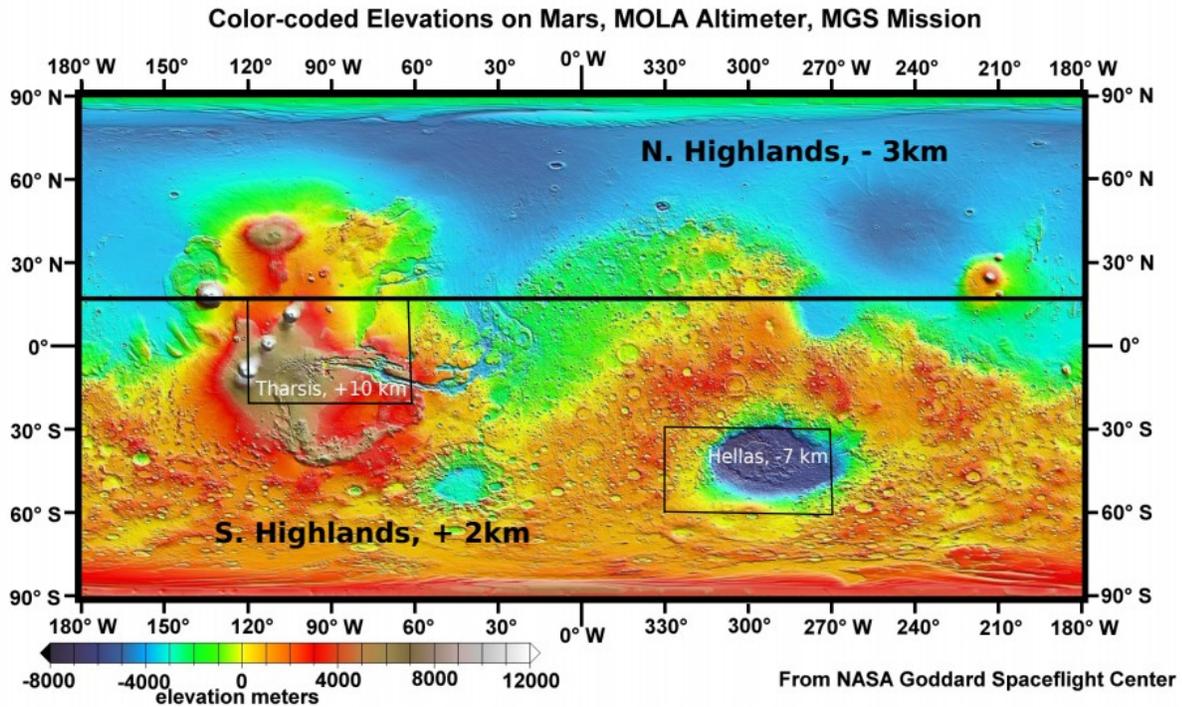

**Figure A1:** Idealized topographical representations for present Mars (Tharsis region, Hellas Basin, Northern Highlands and Southern Highlands) as implemented in Figure 3 overlain on the Mars Orbiter Laser Altimeter (MOLA) map. Image credits: Adapted from the original MOLA map created by the MOLA team at the NASA Goddard Space Flight Center